# *Spatiotemporal Isotropic-to-Anisotropic Meta-Atoms*

*Victor Pacheco-Peña[1] and Nader Engheta[2]*

[1]*School of Mathematics, Statistics and Physics, Newcastle University, Newcastle Upon Tyne, NE1 7RU, United Kingdom*
[2]*Department of Electrical and Systems Engineering, University of Pennsylvania, Philadelphia, PA 19104, USA*
*email: Victor.Pacheco-Pena@newcastle.ac.uk, engheta@ee.upenn.edu

**Metamaterials and metasurfaces are designed by periodically arranged subwavelength geometries, allowing a tailored manipulation of the electromagnetic response of matter. Here, we exploit temporal variations of permittivity inside subwavelength geometries to propose the concept of** *spatiotemporal meta-atoms* **having time-dependent properties. We exploit isotropic-to-anisotropic temporal boundaries within spatially subwavelength regions where their permittivity is rapidly changed in time. In so doing, it is shown how resulting scattered waves travel in directions that are different from the direction of the impinging wave, and depend on the values of the chosen anisotropic permittivity tensor. To provide a full physical insight of their performance, multiple scenarios are studied numerically such as the effect of using different values of permittivity tensor, different geometries of the spatiotemporal meta-atom and time duration of the induced isotropic-to-anisotropic temporal boundary. The intrinsic asymmetric response of the proposed spatiotemporal meta-atoms is also studied demonstrating, both theoretically and numerically, its potential for an at-will manipulation of scattered waves in real time. These results may open new paradigms for controlling wave-matter interactions and may pave the way for the next generation of metamaterials and metasurfaces by unleashing their potential using four-dimensional (4D) unit cells.**

1. **Introduction**

Metamaterials and metasurfaces (as their 2D version) have opened new ways to control and manipulate fields and waves at-will, and have been a hot research topic given their ability to produce artificially engineered media with electromagnetic (EM) properties not easily available in nature[1–4]. They have been proposed and experimentally demonstrated in multiple frequency ranges including radio frequencies, microwave and millimeter up to the optical regime[5–10] and they have been used in groundbreaking applications such as sensors[11–13], quantum devices and technologies [14], antennas and lenses [15–24], beam steerers [25–27], tunable metamaterials and surfaces [28–31], optical circuits[32,33], analogue computing [34–36] and parity-time-symmetric systems[37,38], to name a few.

To design such artificial media, one can exploit subwavelength unit cells (known as meta-atoms) spatially placed in often periodic (and sometimes randomly located) arrangements. In so doing (and considering the materials, geometries and orientation of such subwavelength meta-atoms) it is possible to design artificial EM media having an at-will tailored physical responses (of effective permittivity, $\varepsilon$, and effective permeability, $\mu$) enabling extreme values such as negative or near-zero values of refractive index [39–44].

Since their conception, metamaterials and metasurfaces have been mostly investigated in the frequency domain (time-harmonic scenario) where EM wave propagation can be manipulated by using spatial inhomogeneities (spatial modulation). Recently, the manipulation of EM waves using temporal and spatiotemporal metamaterials has gained a prominent attention in the scientific community, opening new ways to fully manipulate EM radiation both in space and time (*x,y,z,t*). Temporal modulation of media was first studied by Morgenthaler[45] and Fante[46] several decades ago by considering a monochromatic wave traveling within a spatially unbounded medium. It was shown theoretically that, if the relative permittivity of the whole medium is rapidly changed (with a time duration smaller than the period *T* of the incident wave) from $\varepsilon_{r1}$ to $\varepsilon_{r2}$ at t = $t_1$ (both values isotropic and larger than one) a set of forward (FW) and backward (BW) waves can be produced and, interestingly, vector ***k*** is preserved while frequency is changed. Such step-like temporal function of EM parameters of media in time are known as *temporal boundaries* with the FW and BW waves being the temporal analogue of the transmitted and reflected waves, respectively, produced at the spatial interface between two spatially semi-infinite media (spatial boundary)[47].

Temporal and spatiotemporally modulated metamaterials and metasurfaces have become a cutting-edge research field enabling interesting potential applications[48–53] ranging from nonreciprocity [54,55], Fresnel drag [56], temporal gratings[57], effective medium theory[58,59], antireflection temporal coatings[60], frequency conversion[61–63], inverse prism [64], holography[65], temporal aiming[66] and transmission without reflection in time modulated metamaterials using techniques such as our recently proposed temporal Brewster angle [67]. In most of these examples, EM waves are manipulated by considering that the whole medium where the waves travel is modulated in time to induce a temporal boundary. Recently, the interaction of EM waves using externally time-modulated meta-atoms has also been explored showing how metamaterials and metasurfaces can benefit from sinusoidally temporal modulations of subwavelength geometries[68]. However, one may ask: what would happen if isotropic-to-anisotropic temporal boundaries are applied to spatially subwavelength geometries? What interesting physical phenomena can be produced when rapidly changing in time the relative permittivity of a meta-atom from an isotropic value to an anisotropic tensor?

Motivated by the interesting opportunities that temporal and spatiotemporal metamaterials can provide, in this work we propose the effects of temporal boundaries in spatially subwavelength material regions to enable what we call *spatiotemporal isotropic-to-anisotropic meta-atoms*. In a recent work [66] we have demonstrated theoretically and numerically how the direction of energy propagation (*S*) of an obliquely incident *p*-polarized wave can be modified in real time by rapidly changing in time the permittivity of the medium where the wave travels from isotropic $\varepsilon_{r1}$ to an anisotropic permittivity tensor $\overline{\overline{\varepsilon_{r2}}} = \{\varepsilon_{r2z}, \varepsilon_{r2x}\}$ (subscript "*r*" denotes the relative values with respect to that of free space. All values are assumed to be larger than unity. The material resonance frequencies are assumed to be much higher than the operating frequency, i.e., no dispersion is considered). We have also exploited this temporal aiming technique[66] to achieve FW wave propagation and frequency conversion without exciting BW waves; i.e., a technique that we called temporal Brewster angle[67]. Here, we explore such isotropic-to-anisotropic temporal boundaries in two-dimensional (2D) subwavelength geometries rather than in spatially unbounded media. We provide an in-depth study of such spatiotemporal meta-atoms demonstrating both numerically and theoretically how the direction of the resulting "temporal" scattering produced by the isotropic-to-anisotropic temporal boundaries induced in subwavelength geometries can be manipulated at-will by properly engineering the values of the relative permittivity tensor $\overline{\overline{\varepsilon_{r2}}} = \{\varepsilon_{r2z}, \varepsilon_{r2x}\}$. All designs are validated using the time-domain solver of the commercial full-wave simulation software COMSOL Multiphysics®.

## 2. Concept: isotropic to anisotropic temporal boundaries

To begin with, a schematic representation of the proposed spatiotemporal isotropic-to-anisotropic meta-atom is shown in Figure 1. In this work, we consider all media to be non-magnetic, i.e., with $\mu_r = 1$. The background medium is isotropic with a constant (time-invariant) relative permittivity defined as $\varepsilon_{background} = \varepsilon_{r1}$. We consider a single two-dimensional (2D) subwavelength meta-atom being immersed in such spatially unbounded background medium and being illuminated by an obliquely incidence (with incidence angle $\theta_1$) *p*-polarized wave (with the magnetic field polarized along the out-of-plane *y*-axis). All quantities are independent of *y* variable.

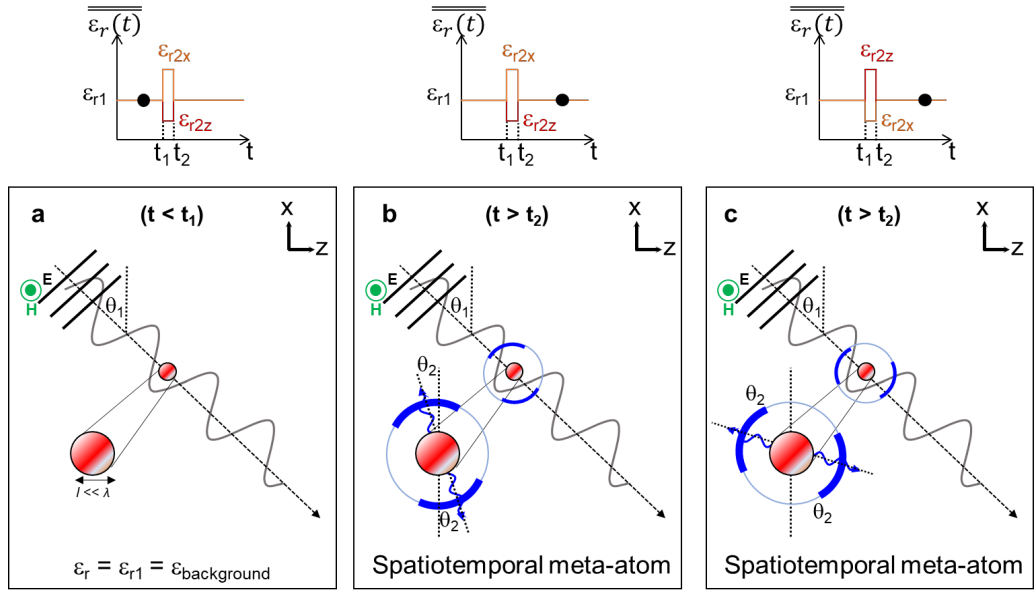

**Figure 1.** Schematic representation of the proposed spatiotemporal isotropic-to-anisotropic two-dimensional (2D) meta-atom (red circles) immersed in a time-independent background medium. The spatiotemporal meta-atom is illuminated by an obliquely incident *p*-polarized monochromatic wave. (a) For times t < $t_1$, the $\varepsilon_r$ of the meta-atom is isotropic and equal to that of the background medium $\varepsilon_{r1}$, therefore there is no scatterer present. In this context, no spatial scattering is produced. (b,c) Schematic representation of the cylindrical wave radiated by the spatiotemporal meta-atoms when their $\varepsilon_r$ is changed in time to an anisotropic tensor $\overline{\overline{\varepsilon_{r2}}}= [\varepsilon_{r2z}, \varepsilon_{r2x}]$ at t = $t_1$ and then returned to isotropic $\varepsilon_{r1}$ at t = $t_2$. The direction of the radiated wave ($\theta_2$) is dependent on the values of the tensor $\overline{\overline{\varepsilon_{r2}}}$ and incidence angle of the monochromatic *p*-polarized wave ($\theta_1$).

As shown in Figure 1(a), for times t < $t_1$, $\varepsilon_r$ of the meta-atom is $\varepsilon_{r1}$; i.e., the same as the background medium, so no scatterer is present before t < t$_1$. Note that we have chosen this value as the same as the background medium to remove any spatial scattering for times before inducing the temporal boundaries (t < t$_1$) in the meta-atoms. Our concept would still be applicable if the scatterer is different from the background medium for t < t$_1$, but here we consider this assumption for the sake of simplicity. At t = t$_1$, the $\varepsilon_r$ of the meta-atoms is changed to an anisotropic permittivity tensor $\overline{\overline{\varepsilon_{r2}}} = \{\varepsilon_{r2z}, \varepsilon_{r2x}\}$. In so doing, the meta-atom will act as a new source creating a cylindrical wave (see Figure 1(b,c) for an schematic representation) which will travel in a direction determined by the induced tensor $\overline{\overline{\varepsilon_{r2}}} = \{\varepsilon_{r2z}, \varepsilon_{r2x}\}$. Importantly, differently to our previous temporal aiming work [66] where the $\varepsilon_r$ of the whole spatially unbounded medium was changed from isotropic-to-anisotropic tensor, here the rapid change of $\varepsilon_r$ is only applied to the subwavelength 2D meta-atom, meaning that the new radiated wave from the meta-atom travel within the time-invariant background medium, enabling the possibility of an arbitrary manipulation of the direction of such "temporal" scattering. Finally, to avoid "spatial" scattering, the $\varepsilon_r$ of the meta-atom is then quickly returned to the initial isotropic value $\varepsilon_{r1}$ at t = t$_2$.

To calculate the angles ($\theta_2$) of the emitted wave produced by the proposed spatiotemporal meta-atom, we can exploit our recent approach for isotropic-to-anisotropic temporal boundaries in spatially unbounded media [66,67]. As we have shown in [66,67], the wave that experiences such temporal boundary will preserve the wavenumber **k** while direction of the energy propagation defined by the Poynting vector (**S**) is modified [($\theta_1 = \theta_{1k,S} = \theta_{2k}$) ≠ ($\theta_2 = \theta_{2S}$)]. This new direction of the energy propagation **S** ($\theta_2$) can be theoretically calculated by considering that the permittivity tensor $\overline{\overline{\varepsilon_{r2}}} = \{\varepsilon_{r2z}, \varepsilon_{r2x}\}$ will modify the amplitude of the $E_x$ and $E_y$ components of the electric field. Hence, $\theta_2$ can be mathematically expressed via the following closed-form equation[66]:

$$\theta_2 = \theta_{2S} = tan^{-1}\left[tan(\theta_1)\left(\frac{\varepsilon_{r2z}}{\varepsilon_{r2x}}\right)\right] \qquad (1)$$

To evaluate the implications of the expression above, the analytical values of $\theta_2$ are shown in Figure 2 considering different incident angles of the monochromatic *p*-polarized wave (namely $\theta_1$ =5°, $\theta_1$ =45° and $\theta_1$ =65°). We consider that $\varepsilon_r$ is changed from $\varepsilon_{r1}$=10 to a tensor $\overline{\overline{\varepsilon_{r2}}} = \{\varepsilon_{r2z}, \varepsilon_{r2x}\}$ at t = t$_1$. As it is shown in Figure 2, $\theta_2$ is strongly dependent on $\theta_1$, $\varepsilon_{r1}$ and $\overline{\overline{\varepsilon_{r2}}}$, as expected from Eq. (1). For instance, it can be observed how larger values of $\theta_2$ can be achieved when increasing the angle of the monochromatic wave before inducing the temporal boundary ($\theta_1$). To guide the eye and for the sake of completeness, we extracted $\theta_2$ from Figure 2(a-c) considering two different fixed values of $\varepsilon_{r2x}$ = 12.5 and $\varepsilon_{r2x}$ = 5 while varying $\varepsilon_{r2z}$. The results are presented in Figure 2(d-f), respectively. From Figure 2(d-f) one can clearly note the influence of both $\theta_1$ and $\overline{\overline{\varepsilon_{r2}}}$ on $\theta_2$. For example, if $\varepsilon_{r2x}$ = 12.5, $\theta_2$ can be modified within the range of [~0.4° to ~8°], [~4.5° to ~58°] or [~9.7° to ~74°] when the $\varepsilon_{r2z}$ is varied from [$\varepsilon_{r2z}$ = 1 to $\varepsilon_{r2z}$ = 20, respectively] considering incident angles of $\theta_1$ = 5°, $\theta_1$ = 45° and $\theta_1$ = 65°, respectively (see black lines in Figure 2(d-f)). The values of $\theta_2$ can then be further increased to be within the range of [~1° to ~19.3°], [~11.3° to ~76°] or [~23.2° to ~83.4°] when using the same range of $\varepsilon_{r2z}$ and values of $\theta_1$, respectively, by simply reducing $\varepsilon_{r2x}$ to $\varepsilon_{r2x}$ = 5 (see blue lines in Figure 2(d-f)).

Note that this approach, which we have proposed and demonstrated numerically in spatially unbounded temporal metamaterials, is also valid for the spatiotemporal meta-atom shown in Figure 1(a) where, as it will be shown in the following sections, the newly emitted wave, resulted from applying the first isotropic-to-anisotropic temporal change of $\varepsilon_{r2}$, will have the same angle defined by Eq. (1). Hence, given the subwavelength size of the meta-atom, the new radiated wave will exit the meta-atom and preserve its direction while traveling within the time-independent background medium, as explained above.

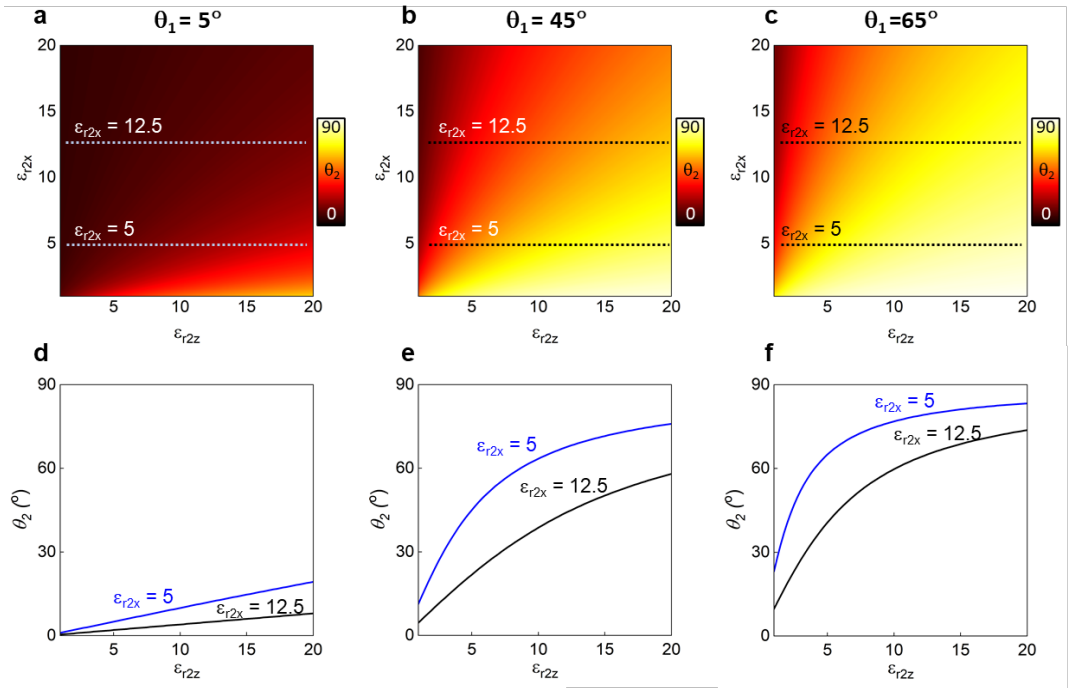

**Figure 2.** Analytically evaluated values of the angle of the energy propagation for times t > t$_1$ ($\theta_2$, calculated using Equation 1) for a monochromatic *p*-polarized wave traveling in a spatially unbounded medium with an incidence angle $\theta_1$. The values of $\theta_2$ are calculated after $\varepsilon_r$ of the spatially unbounded medium is changed from isotropic $\varepsilon_{r1}$ = 10 to a tensor $\overline{\overline{\varepsilon_{r2}}}$ = [$\varepsilon_{r2z}$, $\varepsilon_{r2x}$] at t = t$_1$. (a-c) $\theta_2$ as a function of $\varepsilon_{r2z}$ and $\varepsilon_{r2x}$ for incidence angles $\theta_1$ = 5°, $\theta_1$ = 45° and $\theta_1$ =

65°, respectively. (d-f) $\theta_2$ extracted from the horizontal dotted lines in (a-c), respectively, considering fixed values of $\varepsilon_{r2x}$ = 5 (blue lines) and $\varepsilon_{r2x}$ = 12.5 (black lines) while varying $\varepsilon_{r2z}$.

## 3. Results and discussion

In this section we evaluate the performance of the proposed 2D spatiotemporal meta-atom considering different 2D subwavelength geometries, namely cylinders and squares, of diameter/lateral size of $l \ll \lambda$ ($l = 0.1\lambda$, with $\lambda$ being the wavelength of the incident wave inside the background medium). The relative permittivity of the background medium is again considered to be time invariant with a value of $\varepsilon_{r1}$=10. Without loss of generality, we make use of an obliquely incident monochromatic p-polarized wave with an incidence angle $\theta_1$ = 25° to illuminate the meta-atoms. As in the previous section, all media in this work is considered to be non-magnetic ($\mu_r$= 1). The numerical calculations are carried out using the time-domain solver of the commercial software COMSOL Multiphysics ® with the same setup as in[66].

### 3.1 Tailoring the temporally scattered waves using spatiotemporal meta-atoms.

The analytical results of the angles of the energy propagation after inducing an isotropic-to-anisotropic temporal boundary ($\theta_2$), calculated from Eq. (1), are shown in Figure 3(a) as a function of the relative permittivity tensor $\overline{\overline{\varepsilon_{r2}}} = \{\varepsilon_{r2z}, \varepsilon_{r2x}\}$. For completeness, the values of $\theta_2$ as a function of $\varepsilon_{r2x}$ when fixing $\varepsilon_{r2z}$ to $\varepsilon_{r2z} = 1$, $\varepsilon_{r2z} = 3$, $\varepsilon_{r2z} = 5$, and $\varepsilon_{r2z} = 8$ are shown in Figure 3(b). As observed, similar results as those shown in Figure 2 are obtained where $\theta_2$ is strongly dependent on $\theta_1$ and $\overline{\overline{\varepsilon_{r2}}}$, as expected.

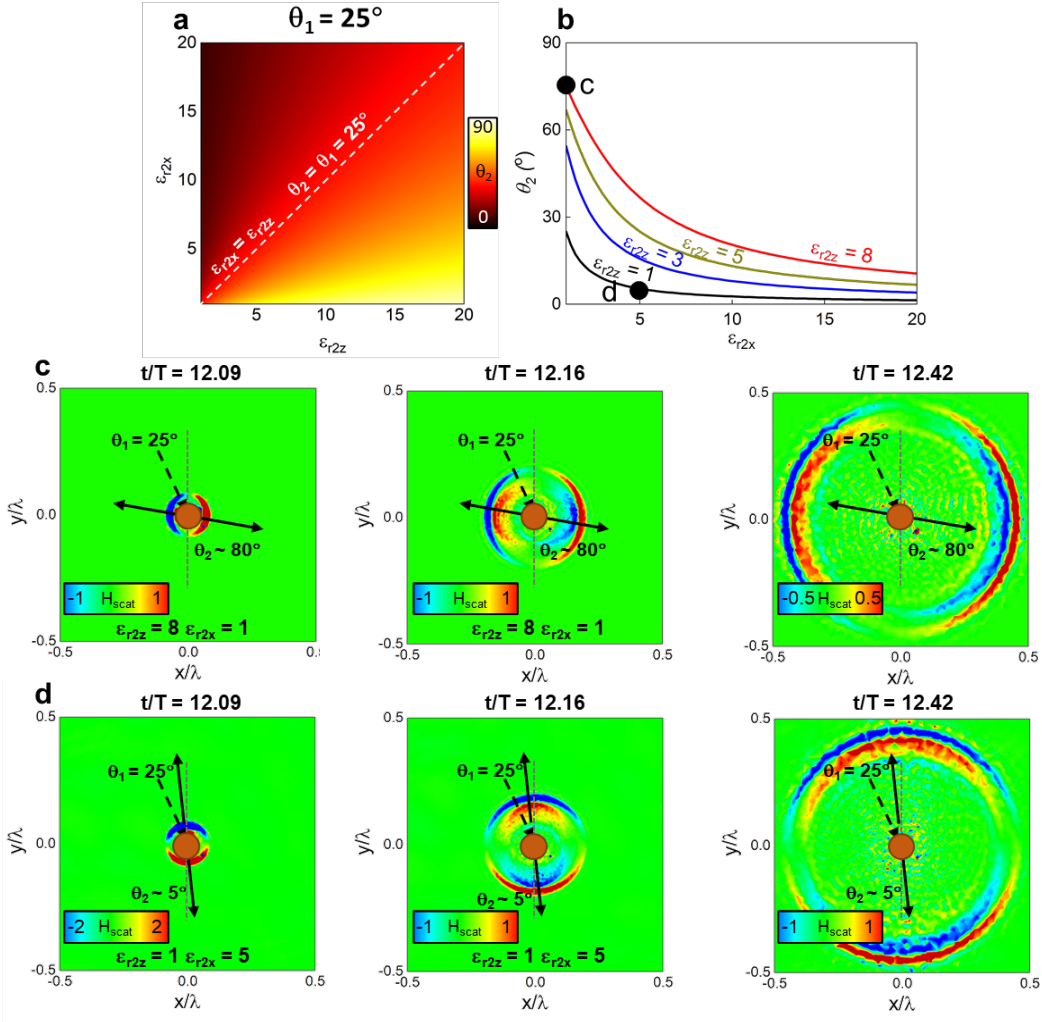

**Figure 3.** (a) $\theta_2$ as a function of tensor $\overline{\overline{\varepsilon_{r2}}} = [\varepsilon_{r2z}, \varepsilon_{r2x}]$ considering an incidence angle of a monochromatic p-polarized wave of $\theta_1 = 25°$, evaluated for a spatially unbounded case. (b) $\theta_2$ as a function of $\varepsilon_{r2x}$ (extracted from panel (a)) for fixed values of $\varepsilon_{r2z}$ namely $\varepsilon_{r2z} = 1$ (black), $\varepsilon_{r2z} = 3$ (blue), $\varepsilon_{r2z} = 5$ (dark yellow) and $\varepsilon_{r2z} = 8$ (red). The black dots in the panel specify the cases shown in panels (c) and (d). Numerical results of the scattered magnetic field distribution $H_{scat}$ at different times $t > t_2$ for cases where $\varepsilon_r$ from isotropic $\varepsilon_{r1} = 10$ to a tensor of (c) $\overline{\overline{\varepsilon_{r2}}} = [\varepsilon_{r2z} = 8, \varepsilon_{r2x} = 1]$ and $\overline{\overline{\varepsilon_{r2}}} = [\varepsilon_{r2z} = 1, \varepsilon_{r2x} = 5]$ at $t = t_1$ and then returned to $\varepsilon_{r1} = 10$ at $t = t_2$. All these results consider $\tau = t_2 - t_1 \sim 0.04T$. In all the simulations, the incident magnetic field $H_{inc}$ has an amplitude of $3 \times 10^{-3}$ A/m. All the color scales are in $10^{-3}$ A/m.

Let us know numerically evaluate the scattered wave being emitted by the proposed spatiotemporal meta-atom, with circular-cylindrical cross section of radius $0.05\lambda$, using different values of the tensor $\overline{\overline{\varepsilon_{r2}}}$. For the numerical simulations, we consider that $\varepsilon_r$ of the meta-atoms is rapidly changed in time from isotropic $\varepsilon_{r1}=10$ to a tensor $\overline{\overline{\varepsilon_{r2}}} = \{\varepsilon_{r2z}, \varepsilon_{r2x}\}$ at a time $t = t_1 = 12.01T$ and it is returned to $\varepsilon_{r1}=10$ at $t = t_2 = 12.05T$. As described in the previous section, note that here $\tau = t_2 - t_1 \ll T$ in order to avoid/lessen spatial scattering (i.e., the scattering that would be produced if the incident monochromatic p polarized wave could interact with an anisotropic subwavelength particle). A discussion about such spatial scattering will be presented later in this manuscript. With this configuration, the numerical results of the scattered magnetic field distribution (calculated as $H_{scat} = H_{total} – H_{inc}$, with $H_{total}$ and $H_{inc}$ being the magnetic field with and without the presence of the spatiotemporal meta-atom, respectively) at different times after inducing the temporal boundaries (t > $t_2$) are shown in Figure 3 (c,d) considering a tensor of $\overline{\overline{\varepsilon_{r2}}} = \{\varepsilon_{r2z} = 8, \varepsilon_{r2x} = 1\}$ and $\overline{\overline{\varepsilon_{r2}}} = \{\varepsilon_{r2z} = 1, \varepsilon_{r2x} = 5\}$, respectively.

From these results, one can clearly note how the scattered field (emitted wave) traveling within the (time-independent) background medium has an angle of $\theta_2 \sim 80°$ and $\theta_2 \sim 5°$ for each case of $\overline{\overline{\varepsilon_{r2}}}$ in Figure 2(c,d), respectively, in line with the analytical values from Figure 2(a) which predicts $\theta_2 = 75°$ and $\theta_2 = 5.32°$, respectively. Importantly, note that the analytically predicted values consider the case when the $\varepsilon_r$ is changed from isotropic-to-anisotropic and kept to this value while the numerical simulations consider that $\varepsilon_r$ is returned to isotropic at t = $t_2$. As discussed before, when applying the first temporal boundary at t = $t_1$, the spatiotemporal meta-atom will emit a cylindrical wave with angle $\theta_2$. Given the subwavelength size of the meta-atom, this scattered wave almost completely leaves the meta-atom and propagates in the background medium with an angle being approximately equals to $\theta_2$. This has interesting implications as the proposed spatiotemporal meta-atom could potentially be engineered to emit such scattered waves at any desired output angle by simply modifying the values of the tensor $\overline{\overline{\varepsilon_{r2}}}$, a mechanism that can open new directions for EM wave manipulation using 4D metamaterials.

**3.2 Changing the geometry of the spatiotemporal meta-atom.**

In section 3.1 we discussed the scattered wave produced by 2D circular-cylindrical spatiotemporal meta-atoms demonstrating how its direction can be tailored by using isotropic-to-isotropic temporal boundaries with properly chosen values for the tensor $\overline{\overline{\varepsilon_{r2}}}$. However, what would happen if the shape of the meta-atom is different? Intuitively, one can foresee that the shape of the meta-atom would not be an important parameter for the resulting angle of the scattered wave ($\theta_2$) given its subwavelength size ($l << \lambda$). In the following study, this question will be addressed by considering 2D square (with a lateral size of 0.1λ) and circular (with a diameter of 0.1λ) shapes for the spatiotemporal meta-atom.

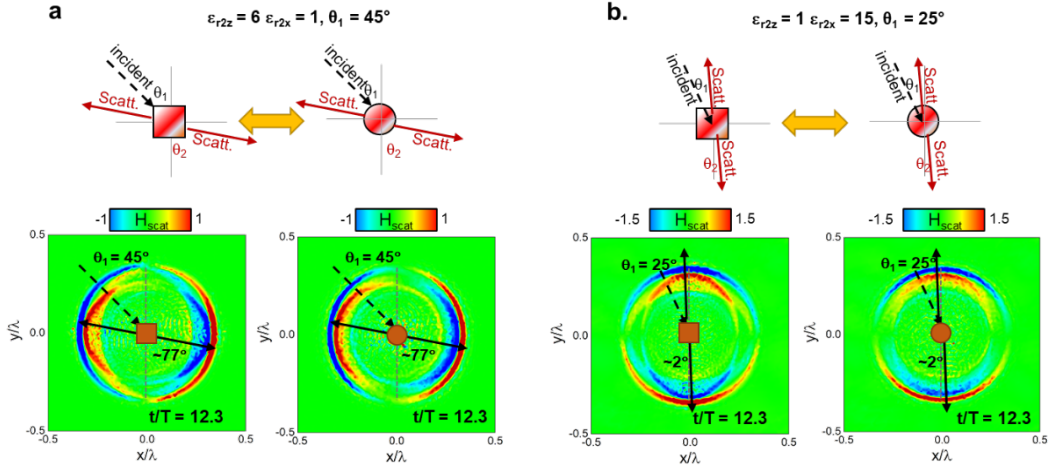

**Figure 4.** Numerical simulations for the scattered field distributions at a time t = 12.3T (i.e., t > $t_2$) for spatiotemporal meta-atoms with square and cylindrical spatial shapes with lateral/diameter size of 0.1λ using an incident angle of (a) $\theta_1 = 45°$ and (b) $\theta_1 = 25°$. The $\varepsilon_r$ of the meta-atoms is changed from isotropic $\varepsilon_{r1} = 10$ to a tensor of (a) $\overline{\overline{\varepsilon_{r2}}} = [\varepsilon_{r2z} = 6, \varepsilon_{r2x} = 1]$ and (b) $\overline{\overline{\varepsilon_{r2}}} = [\varepsilon_{r2z} = 1, \varepsilon_{r2x} = 15]$ at t = $t_1$ and then returned to $\varepsilon_{r1} = 10$ at t = $t_2$. All these results consider $\tau = t_2 - t_1 \sim 0.04T$. In all the simulations, the incident magnetic field $H_{inc}$ has an amplitude of $3 \times 10^{-3}$ A/m. All the color scales are in $10^{-3}$ A/m.

Here the spatiotemporal meta-atoms are illuminated with a monochromatic obliquely incident *p*-polarized wave with an incidence angle of $\theta_1 = 45°$ or $\theta_1 = 25°$ (see Figure 4(a) and Figure 4(b), respectively). The numerical results of the scattered magnetic field distribution for the squared and cylindrical meta-atoms after inducing the temporal boundaries (t > $t_2$) are shown in Figure 4(a,b) considering values of the tensor $\overline{\overline{\varepsilon_{r2}}}$ of $\overline{\overline{\varepsilon_{r2}}} = \{\varepsilon_{r2z} = 6, \varepsilon_{r2x} = 1\}$ and $\overline{\overline{\varepsilon_{r2}}} = \{\varepsilon_{r2z} = 1, \varepsilon_{r2x} = 15\}$, respectively. By comparing the results for both 2D shapes of the spatiotemporal meta-atoms, one can clearly notice that the direction of the scattered wave $\theta_2$ is effectively the same, as expected due to the subwavelength size of the meta-atoms. Finally, note that the resulting values of $\theta_2$ are $\theta_2 \sim 77°$ and $\theta_2 \sim 2°$ when using $\overline{\overline{\varepsilon_{r2}}} = \{\varepsilon_{r2z} = 6, \varepsilon_{r2x} = 1\}$ and $\overline{\overline{\varepsilon_{r2}}} = \{\varepsilon_{r2z} = 1, \varepsilon_{r2x} = 15\}$, respectively, in agreement with Equation 1 which theoretically predict values of $\theta_2 = 80.54°$ and $\theta_2 = 1.78°$, respectively.

### 3.3 Asymmetric response of the spatiotemporal meta-atoms

Based on the results shown in Figure 2-4, one may ask the following question: does this spatiotemporal meta-atom has an intrinsically nonreciprocal response? We have discussed how the direction of the scattered wave produced by the spatiotemporal meta-atom can be tailored by changing the tensor $\overline{\overline{\varepsilon_{r2}}}$ and incident angle $\theta_1$. Hence, one may foresee a nonreciprocal behavior of the meta-atom.

To study this, let us consider the case shown in Figure 5(a) using a 2D cylindrical shape (diameter of 0.1λ) of the meta-atom. As in the previous section, an incident monochromatic *p*-polarized wave with $\theta_1 = 25°$ is again traveling from top-to-bottom on the *xy* plane. The $\varepsilon_r$ of the meta-atom is isotropic with $\varepsilon_{r1} = 10$ for t < t$_1$, it is changed to an anisotropic tensor $\overline{\overline{\varepsilon_{r2}}} = \{\varepsilon_{r2z} = 5, \varepsilon_{r2x} = 2\}$ at t = t$_1$ and returned to $\varepsilon_{r1} = 10$ at t = t$_2$ (with τ = 0.04T). With this setup, the numerical results of the incident magnetic field distribution ($H_{inc}$) at a time t < t$_1$ is shown in Figure 5(d) along with the scattered magnetic field distribution ($H_{scat}$) calculated at different times t > t$_2$ (second and third panels). From these results, the direction of the scattered wave produced by the spatiotemporal meta-atom is $\theta_2 \sim 49°$, in agreement with the theoretically predicted value from Equation 1 of $\theta_2 = 49.38°$. To evaluate the possibility of using the proposed spatiotemporal meta-atom as a nonreciprocal subwavelength particle, we can then use the angle $\theta_2 \sim 49°$ as the new incidence angle of the incident monochromatic wave and illuminate the meta-atom from bottom-to-top with an angle of $\theta'_1 = 49°$. With this configuration, the instantaneous incident ($H_{inc}$) and scattered ($H_{scat}$) magnetic field distribution calculated at the same time as the result discussed in Figure 5(d) are shown in Figure 5(e), respectively. From these results, once can clearly see how the resulting scattered wave produced by the spatiotemporal meta-atom is directed to an angle $\theta'_2 \sim 71°$ (again in agreement with Equation 1 which predicts $\theta'_2 \sim 70.8°$), i.e., with $\theta'_2 \neq \theta_1$, the scattered field is not directed towards 25° (incidence angle used for Figure 5(d)), a value that would account for a reciprocal behavior of the meta-atom.

These results demonstrate how spatiotemporal meta-atoms that are modulated in time with isotropic-to-anisotropic functions of $\varepsilon_r$ can expectedly exhibit an intrinsic nonreciprocal behavior and can be used to steer the generated scattered wave in real time, features that could open the way towards new paradigms for spatiotemporal control of EM wave propagation using 4D metamaterials.

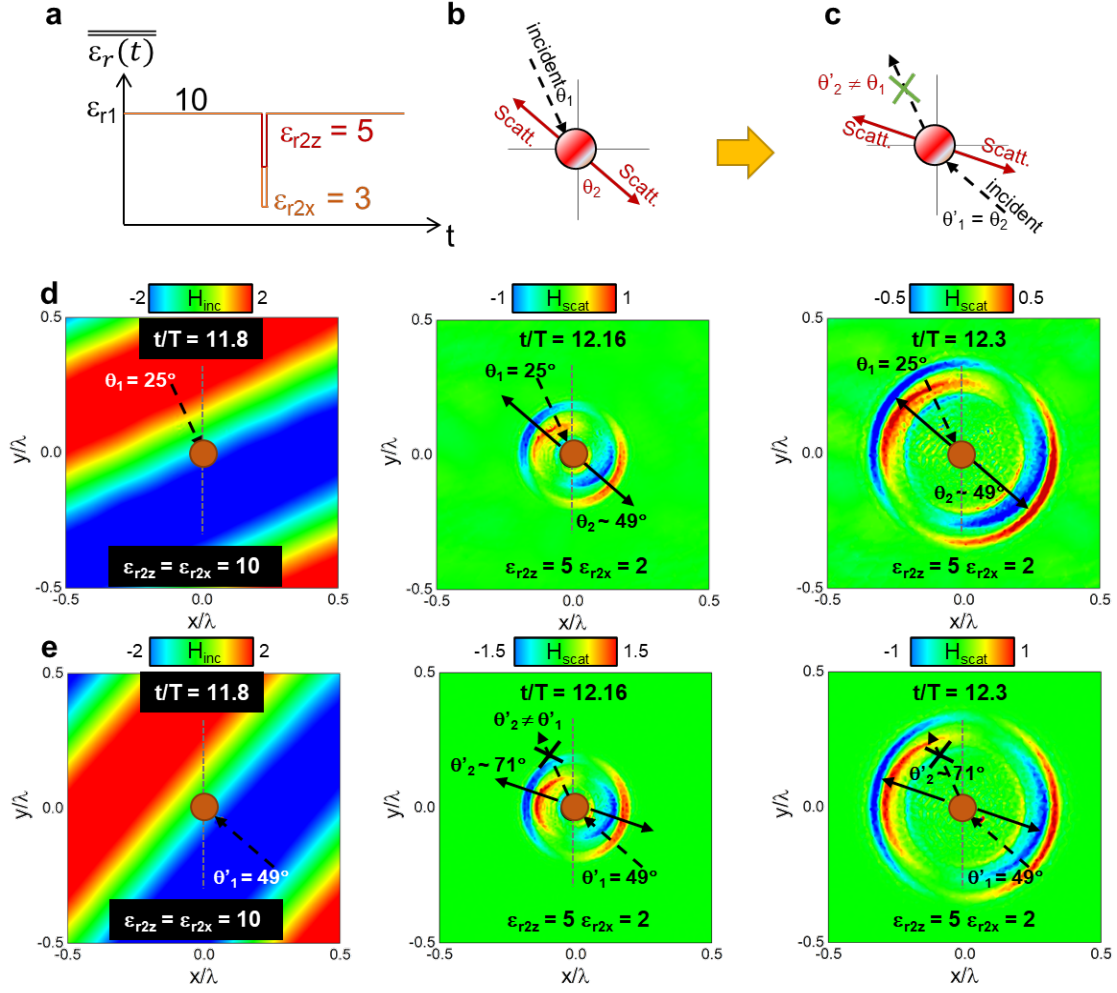

**Figure 5.** Nonreciprocal response of the spatiotemporal meta-atom. (a) Temporal variation of $\varepsilon_r$ of the meta-atom. (b) schematic representation of the incident wave traveling along $\theta_1 = 25°$ (from top-to-bottom) showing the direction of the scattered wave that would be produced by the spatiotemporal meta-atom after $t > t_2$ when the meta-atom is modulated in time to have the $\varepsilon_r$ as shown in panel (a). (c) schematic representation of intrinsic nonreciprocal response of the spatiotemporal meta-atom depicting the incident wave traveling along $\theta'_1 = \theta_2 = 49°$ (from bottom-to-top) showing the direction of the scattered wave that would be produced by the spatiotemporal meta-atom after $t > t_2$ when the meta-atom is modulated in time to have the $\varepsilon_r$ as shown in panel (a). (d,e) Numerical results of the incident (left panel) and scattered (middle and right panels) magnetic field distributions at different times for the cases depicted in panels (b) and (c), respectively. All these results consider $\tau = t_2 - t_1 \sim 0.04T$. In all the simulations, the incident magnetic field $H_{inc}$ has an amplitude of $3\times10^{-3}$ A/m. All the color scales are in $10^{-3}$ A/m.

### 3.4 Effect of changing $\tau = t_2 - t_1$

All the results presented in the previous sections have been obtained considering $\tau = t_2 - t_1 \sim 0.04T$. Here we will discuss the effect of this parameter on the scattered wave produced by the spatiotemporal meta-atom with induced temporal boundaries. Here we will use two different values of $\tau$, namely $\tau = 0.1T$ and $\tau = T$ (both larger than those used in the previous sections). The temporal function of $\varepsilon_r(t)$ for the spatio-temporal meta-atoms is schematically shown in Figure 6(a) and Figure 6(f), respectively. Let us first examine the case with $\tau = 0.1T$. The numerical results of the scattered magnetic field distribution at different times are shown in Figure 6(b-e). As observed, we have chosen times at t = $t_1^+$, $t_1 < t < t_2$, $t = t_2^-$ and $t > t_2$, respectively. From these results, one can notice how the resulting scattered wave resembles that shown in Figure 5(d). However, note that here the effect of the second temporal boundary at $t = t_2$ starts to become more evident (see Figure 6(e)) with the scattered wave having more oscillations that that shown in Figure 5(d). However, note that spatial scattering in this case is almost negligible as the duration of $\tau \ll T$, as expected.

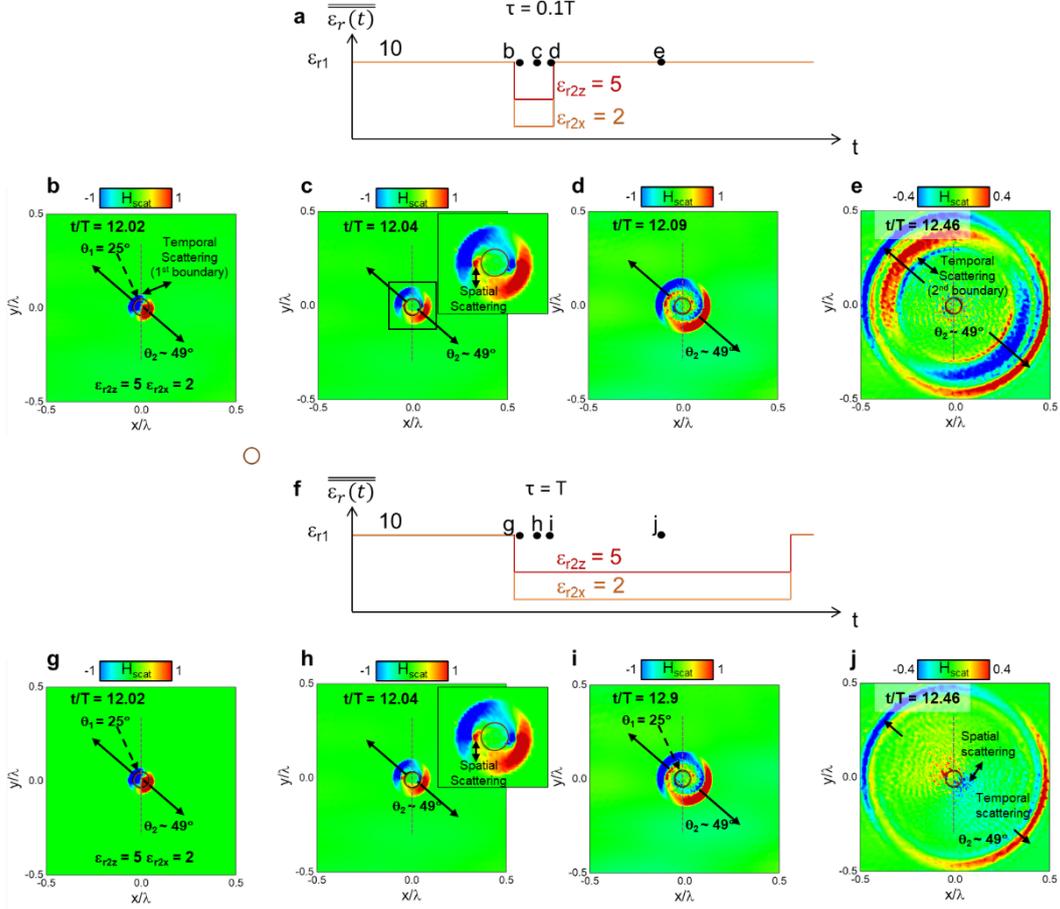

**Figure 6.** Effect of temporal duration $\tau = t_2 - t_1$: (a,f) Temporal variation of $\varepsilon_r$ of the 2D cylindrical meta-atom considering a value of $\tau = 0.1T$ and $\tau = T$ (b-e) and (g-j) Scattered magnetic field at different times for the temporal variation of $\tau = 0.1T$ and $\tau = T$, respectively. In all these results an incident angle of $\theta_I = 25°$ (from top-to-bottom) is considered. In all the simulations, the incident magnetic field $H_{inc}$ has an amplitude of $3 \times 10^{-3}$ A/m. All the color scales are in $10^{-3}$ A/m.

What would happen if we increase the duration of $\tau$? The numerical results of the magnetic field distribution for the case with $\tau = T$ are shown in Figure 6 (g-j) calculated at the same times as in Figure 6(b-e), respectively. Interestingly, by comparing the results shown in Figure 6(d) and Figure 6(i), one can note how the scattered wave produced by the first temporal boundary is present in both cases along with a small spatial scattering (see also Figure 6(c) and Figure 6(h)), as expected.

*13*

Moreover, for a time t = 12.46T, two main differences can be noticed by comparing the results with τ = 0.1T (Figure 6(e)) and τ = T (Figure 6(j)). One is that the number of oscillations of the scattered wave is smaller for the case with τ = T, which is a direct consequence of only one temporal boundary being induced at that time while two temporal boundaries have been induced for the case with τ = 0.1T. The second difference is that spatial scattering is observable for the case with τ = T which is not noticeably present for τ = 0.1T. Again, this is an expected result given that at a time t = 12.46T, the $\varepsilon_r$ of the meta-atom is still anisotropic for the case with τ = T while it is isotropic and equals to the background medium for the case with τ = 0.1T. These results demonstrate how mainly temporal or both spatial and temporal scattering can be produced depending on the values of τ in spatiotemporal meta-atoms with induced isotropic-to-anisotropic temporal boundaries.

4. **Conclusions**

In conclusion, we have shown and studied, both numerically and theoretically, the use isotropic-to-anisotropic temporal changes of permittivity within spatially subwavelength regions rather than in spatially unbounded media. In this context, we have proposed what we called *spatiotemporal isotropic-to-anisotropic meta-atoms*. Several configurations have been presented such as different values of permittivity tensor and shape of the subwavelength particle. We have shown how the direction of the scattered wave produced by the meta-atom when inducing such temporal changes of permittivity can be manipulated at will by carefully crafting the values of the permittivity tensor of the meta-atom. A discussion about the intrinsically non-reciprocal response of the spatiotemporal meta-atom was presented along with the effect of the time duration (τ = $t_2 - t_1$) of the temporal boundary, showing how either only temporal or both spatial and temporal scattering can be achieved by using different values of τ. These results may open new avenues for exploration and exploitation of the next generation of metamaterials and metasurfaces by using 4D spatiotemporally modulated meta-atoms.

**Acknowledgements**

V.P.-P. acknowledges support from the Newcastle University (Newcastle University Research Fellowship). N.E. would like to acknowledge the partial support from the Vannevar Bush Faculty

Fellowship program sponsored by the Basic Research Office of the Assistant Secretary of Defense for Research and Engineering, funded by the Office of Naval Research through grant N00014-16-1-2029.